\documentclass[prl,aps,twocolumn,reprint,amsmath,amssymb]{revtex4}
\usepackage{graphicx}  
\usepackage{dcolumn}   
\usepackage{bm}        
\usepackage{amssymb}   
\usepackage{amsmath}   
\usepackage{chemarrow}
\usepackage{color}
\usepackage{float}
\begin{document}


\title{Non-Gaussian Precision Metrology via Driving through Quantum Phase Transitions}

\author{Jiahao Huang$^{1}$}

\author{Min Zhuang$^{1,2}$}

\author{Chaohong Lee$^{1,2,3}$}
\altaffiliation{Email: lichaoh2@mail.sysu.edu.cn, chleecn@gmail.com}

\affiliation{$^{1}$Laboratory of Quantum Engineering and Quantum Metrology, School of Physics and Astronomy, Sun Yat-Sen University (Zhuhai Campus), Zhuhai 519082, China}

\affiliation{$^{2}$State Key Laboratory of Optoelectronic Materials and Technologies, Sun Yat-Sen University (Guangzhou Campus), Guangzhou 510275, China}

\affiliation{$^{3}$Synergetic Innovation Center for Quantum Effects and Applications, Hunan Normal University, Changsha 410081, China}

\date{\today}

\begin{abstract}
We propose a scheme to realize high-precision quantum interferometry with entangled non-Gaussian states by driving the system through quantum phase transitions.
The beam splitting, in which an initial non-degenerate groundstate evolves into a highly entangled state, is achieved by adiabatically driving the system from a non-degenerate regime to a degenerate one.
Inversely, the beam recombination, in which the output state after interrogation becomes gradually disentangled, is accomplished by adiabatically driving the system from the degenerate regime to the non-degenerate one.
The phase shift, which is accumulated in the interrogation process, can then be easily inferred via population measurement.
We apply our scheme to Bose condensed atoms and trapped ions, and find that Heisenberg-limited precision scalings can be approached.
Our proposed scheme does not require single-particle resolved detection and is within the reach of current experiment techniques.

\end{abstract}


\maketitle
Recent breakthroughs in generating many-body quantum entangled states boost tremendous advances in quantum metrology~\cite{Giovannetti2004, Giovannetti2006, Giovannetti2011, Ludlow2015, Pezze2016}.
Most protocols focus on using Gaussian spin squeezed states.
The spin squeezed states are often created via dynamical evolution with nonlinear interactions, such as spin-twisting~\cite{Kitagawa1993, Gross2010, Riedel2010,  Ma2011, Muessel2015}.
Remarkably, entangled non-Gaussian states (ENGSs) of massive particles such as GHZ states, set a benchmark for approaching the Heisenberg limit in metrology, which offer better scalings than conventional spin squeezed states~\cite{Huang2014}.
These ENGSs can also be generated by dynamical evolution in ultracold atomic gases~\cite{Lucke2011, Bookjans2011, Strobel2014, Gabbrielli2015, Helm2017}, trapped ions~\cite{Monz2011} and optical systems~\cite{Pan2012}.
However, this method requires long evolution time, and the dynamically generated states are always not steady states so that the system parameters must be precisely controlled.
An alternative way for preparing ENGSs is to drive the system through quantum phase transitions (QPTs)~\cite{Lee2006, Lee2009, Zhang2013, Huang2015, Xing2016}.
The generation of entangled states via adiabatic driving is deterministic and more robust, which has been realized in recent experiments~\cite{Luo2017}.

Even though ENGSs can be prepared experimentally, they are always difficult to detect.
To fully exploit their quantum resources for achieving precision beyond standard quantum limit, detectors of single-particle resolution are always needed for reading out entangled sensor states~\cite{Zhang2012, Hume2013}.
Therefore it is important to consider whether one can replace single-particle resolved detection with low-resolution detection (such as population detection).
With the input entangled states generated by nonlinear dynamical evolution, the echo protocols with time-reversal nonlinear dynamics have been theoretically proposed~\cite{Huelga1997, Frowis2016, Davis2016, Macri2016, Szigeti2017, Nolan2017, Fang2017} and experimentally demonstrated~\cite{Linnemann2016, Hosten2016}.
While the driving through QPTs has been proposed for deterministic generation of ENGSs~\cite{Lee2006, Xing2016, Helm2017}, to fully utilize ENGSs for quantum sensing without single-particle resolved detection, can one use the reversely driving through QPTs for beam recombination?

In the Letter, we propose how to implement Heisenberg-limited quantum phase estimation via driving through QPTs between non-degenerate and degenerate regimes.
In our proposal, the beam splitting and recombination are achieved by adiabatically sweeping the system parameter forwardly and inversely through QPTs, respectively.
By sweeping an interacting many-body quantum system from a non-degenerate regime to a degenerate one, its final state always appears as an ENGS.
In the phase accumulation process, the state acquires a phase to be measured.
To extract the phase, the beam recombination, a reversed sweeping from the degenerate regime to the non-degenerate one, is introduced to disentangle the sensor state.
To show the validity of our scheme, as two examples, we apply it to a Bose-Josephson model (for Bose condensed atoms) and a transverse-field Ising model (for trapped ions), and find that the measurement precisions indeed approach the Heisenberg limit.
Since the final states before measurement can be resolved with coarsened detectors, our scheme relaxes the single-particle resolution typically requested in previous schemes via parity measurement~\cite{Campos2003, Anisimov2010, Gerry2010, Huang2015, LuoC2017}.
%

\begin{figure}[!t]
\includegraphics[width=1\columnwidth]{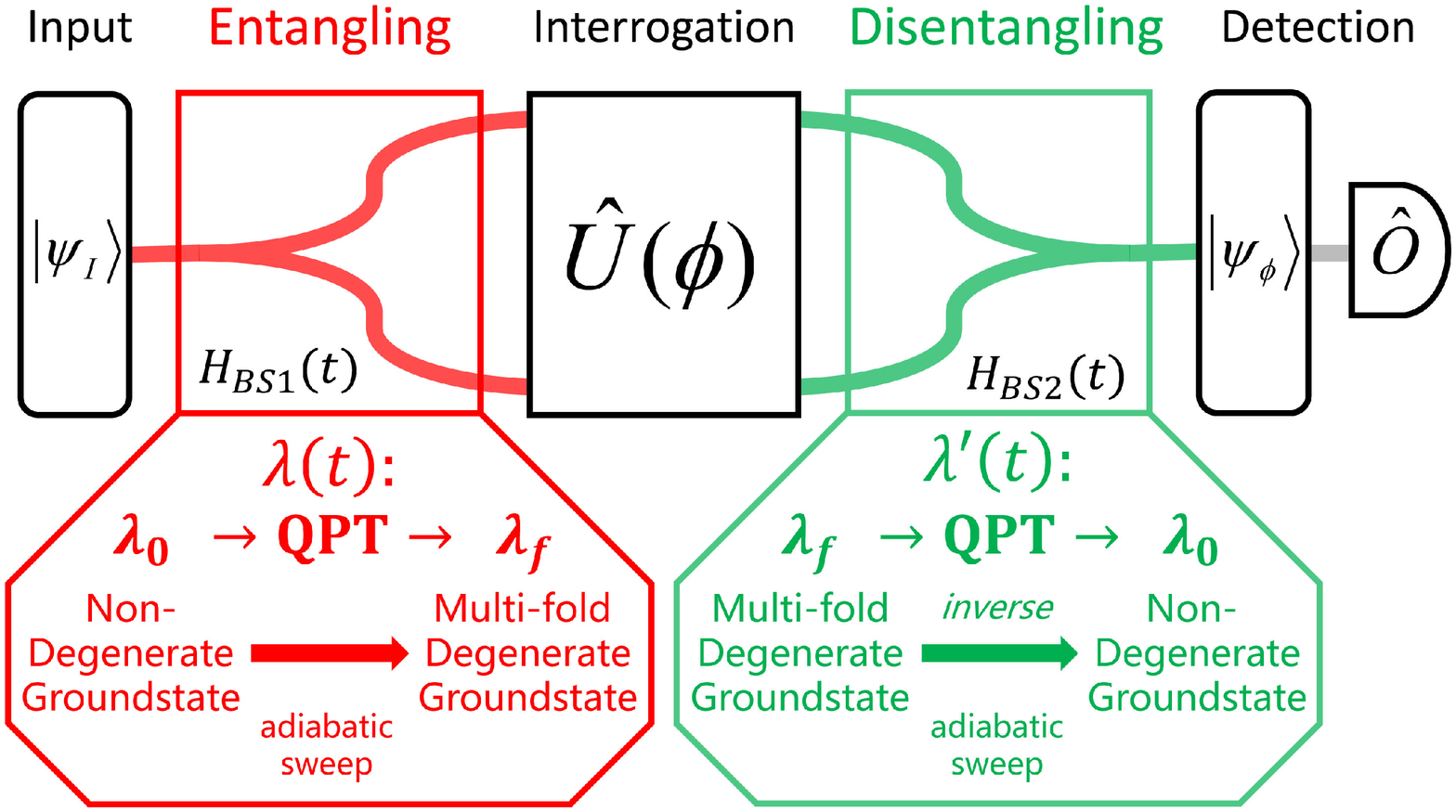}
\caption{\label{Fig1}(color online). Schematic diagram of quantum phase estimation via driving through QPTs. The two beam splitters are achieved by driving $\lambda(t)=R_1(t)/R_2(t)$ forwardly and inversely across the critical point $\lambda_c$. In the first beam splitter, $H_{BS1}(t)=R_1(t) H_1 + R_2(t) H_2$, $\lambda$  adiabatically sweeps from $\lambda_0$ to $\lambda_f$ and the state evolves from a non-degenerate groundstate [$R_1(0) H_1$ dominates] to a degenerate groundstate [$R_2(\tau) H_2$ dominates]. Then the state accumulates the phase $\phi$ in the interrogation process. While in the second beam splitter, $H_{BS2}(t)=R_1'(t) H_1 + R_2'(t) H_2$, $\lambda'$ inversely sweeps from $\lambda_f$ [$R_2(\tau) H_2$ dominates] to $\lambda_0$ [$R_1(0) H_1$ dominates] and the state is gradually disentangled. }
\end{figure}

Our protocol is shown in Fig.~\ref{Fig1}.
We assume all time-evolution dynamics are unitary and set $\hbar=1$.
First, an initial state $|\psi_I\rangle$ is prepared as the groundstate $|\psi\rangle_I$ of $H_{BS1}(0)$.
Then, an entangled state $|\psi_E\rangle$ is gradually created in the beam splitting process $H_{BS1}(t)$ via driving through QPTs.
In the interrogation process, the state acquires a phase $\phi$ via a phase-imprinting evolution $\hat U(\phi)$, that is, $|\psi(\phi)\rangle=\hat U(\phi)|\psi\rangle_E$.
To extract the accumulated phase $\phi$, the beam recombination $H_{BS2}(t)$ and a subsequent measurement of the observable $\hat O$ are implemented.
Here the beam recombination $H_{BS2}(t)$ is achieved by the reversed sweeping of the beam splitting $H_{BS1}(t)$.

\textit{Beam Splitting and Recombination via driving through QPTs.---}
The beam splitting for generating ENGSs can be described by a time-dependent Hamiltonian,
\begin{equation}\label{Ham_QPT}
    H_{BS1}(t) = R_1(t) H_1 + R_2(t) H_2.
\end{equation}
Here, the time-varying parameters $R_1(t)$ and $R_2(t)$ interpolates the Hamiltonians $H_1$ and $H_2$.
We assume the whole duration of beam splitting is $\tau$.
We choose proper Hamiltonians such that the groundstate of $R_1(0) H_1$ is non-degenerate while the groundstate of $R_2(\tau) H_2$ is multi-fold degenerate.
There exists a QPT at the critical point $\lambda_c$, that is, $R_1(t) H_1$ dominates the system when $|\lambda(t)|>|\lambda_c|$ and $R_2(t) H_2$ dominates the system when $0\le |\lambda(t)|<|\lambda_c|$.
Starting from the non-degenerate groundstate of $R_1(0) H_1 + R_2(0) H_2$ ($\lambda_0 \equiv R_1(0)/R_2(0)$ and $ |\lambda_0| > |\lambda_c|$), an entangled groundstate [a specific superposition of the degenerate groundstates of $R_1(\tau) H_1 + R_2(\tau) H_2$] can be obtained with high fidelity if $\lambda$ is adiabatically swept from $\lambda_0$ to $\lambda_f$ ($0 \le |\lambda_f| < |\lambda_c|$).

While the adiabatic operation of $H_{BS1}(t)$ serves as the first beam splitter and generates the entangled input state $|\psi\rangle_E$, an inverse operation of $H_{BS1}(t)$ onto $|\psi\rangle_E$ would in principle disentangle it back to the initial state $|\psi\rangle_I$, which can act as the second beam splitter for recombination.
The second beam splitter can be described by a reversed time-dependent Hamiltonian,
\begin{eqnarray}\label{Ham_QPT2}
    H_{BS2}(t) &=& R_1'(t) H_1 + R_2'(t) H_2, \nonumber \\
    &=& R_1(\tau-t) H_1 + R_2(\tau-t) H_2,
\end{eqnarray}
where $\lambda'(t)=R_1'(t)/R_2'(t)$ is inversely swept from $\lambda_f$ towards $\lambda_0$ with opposite sweeping rate for $H_{BS1}(t)$.

Thus, the state before detection is expressed as $|\psi(\phi)\rangle_R = e^{-i \int_{0}^{\tau'} H_{BS2}(t) dt } \hat U(\phi) e^{-i \int_0^{\tau} H_{BS1}(t) dt} |\psi_I\rangle$ with the splitting duration $\tau$ and the recombination duration $\tau'$.
%
%
Ideally for $\phi=0$, the state $|\psi(\phi)\rangle_R$ is identical to the initial state $|\psi_I\rangle$ (there may exist a globally relative phase) if $\lambda'(\tau') = \lambda_0$.
When $\phi \neq 0$, the nonzero phase will bias the state $|\psi(\phi)\rangle_R$ with respect to the initial state $|\psi_I\rangle$, and one can detect a $\phi$-dependent observable $\langle \hat O\rangle$ to extract the phase information.
In practice, the recombination duration can be shorter than the splitting duration, i.e., $\tau' \le \tau$.
Thus in the inverse sweeping, it is not required to reach $\lambda_0$ exactly and there may exist several optimal values $\lambda'(\tau')=\lambda_{opt}'$ that can achieve the best measurement precision.
Obviously, the recombination via inverse sweeping adds no additional complexity of experimental manipulation.

\textit{Bose-Josephson model.---}
We first illustrate our protocol in the Bose-Josephson model.
Generally, the symmetric Bose-Josephson Hamiltonian reads~\cite{Gross2010, Riedel2010, Strobel2014, Lee2006,  Ribeiro2007}
\begin{equation}\label{Ham_BJ}
    H_{BJ} = H_{\Omega} + H_{\chi} = -\Omega \hat J_x + \frac{\chi}{N} \hat J_z^2,
\end{equation}
with the Josephson coupling strength $\Omega$, the nonlinear atom-atom interaction $\chi$, and the collective spin operators: $\hat J_x = {1\over2} \left(\hat a \hat b^{\dagger} + \hat a^{\dagger} \hat b\right)$, $\hat J_y = {1\over2i} \left(\hat a \hat b^{\dagger} - \hat a^{\dagger} \hat b\right)$, $\hat J_z = {1\over2}\left(\hat b^{\dagger}\hat b - \hat a^{\dagger} \hat a\right)$.
Here $\hat a $ and $\hat b$ are annihilation operators for particles in modes $|a\rangle$ and $|b\rangle$, respectively.
%
%
There exists a transition between non-degenerate and degenerate groundstates when the nonlinearity is negative (i.e. $\chi<0$)~\cite{Lee2006, Lee2009}.
For an $N$-atom system with $\chi<0$, in strong coupling limit ($\Omega \gg |\chi|$), the groundstate is a SU(2) spin coherent state.
%
When $\Omega \rightarrow 0$, the interaction dominates and the two lowest eigenstates become degenerate.
%
%
The transition from non-degeneracy to degeneracy happens at the critical point $\Omega_c/|\chi|=1$, which corresponds to a classical Hopf bifurcation from single-stability to bistability.
Starting from the groundstate of $H_{BJ}$ with large $\Omega$ and adiabatically decreasing $\Omega$ across the critical point $\Omega_c= |\chi|$, one can get a superposition of two self-trapping states, which can be approximately regarded as a spin cat state~\cite{Huang2015}.
%
%
%
In our calculation, we set $\chi=-1$ and thus $\Omega_c=1$.

To shorten the duration, we sweep the Josephson coupling strength $\Omega$ in two stages with different sweeping rates (it is also efficient that the sweeping rate becomes time-dependent and is determined according to the energy spectra~\cite{Peng2014, Luo2016}).
When $\Omega(t) > 1$, the gap between the groundstate and the first excited state is large and so that the sweeping can be faster.
While $\Omega(t) \le 1$, the groundstate and first excited state become nearly degenerate and so that the sweeping should be much slower.
Thus, the sweeping process can be described by
\begin{equation}\label{Omega1}
\Omega(t) =
\begin{cases}
\begin{split}
\Omega_0 - \beta_1 t, \quad \quad \quad 0 \le t \le \tau_c,\\
\Omega_c - \beta_2 (t-\tau_c),   \tau_c < t \le \tau,
\end{split}
\end{cases}
\end{equation}
with the initial Josephson coupling strength $\Omega_0$, $\tau_c={(\Omega_0-\Omega_c)}/{\beta_1}$ and $\tau=\tau_c+{(\Omega_c-\Omega_f})/{\beta_2}$.
Here, $\beta_1$ and $\beta_2$ denote the sweeping rates in the first stage from $\Omega_0$ to $\Omega_c$ and the second stage from $\Omega_c$ to $\Omega_f$ ($0 \le \Omega_f < \Omega_c$), respectively.
%
%
The sweeping ends at different values of $\Omega_f$ would result in different ground states $|\Psi(\Omega_f)\rangle_G$.
We prepare four different input states $|\Psi(\Omega_f)\rangle_G$ for $\Omega_f=(0, 0.25, 0.5, 0.75)$ and analyze their interferometric performances.

\begin{figure}[!t]
\includegraphics[width=1\columnwidth]{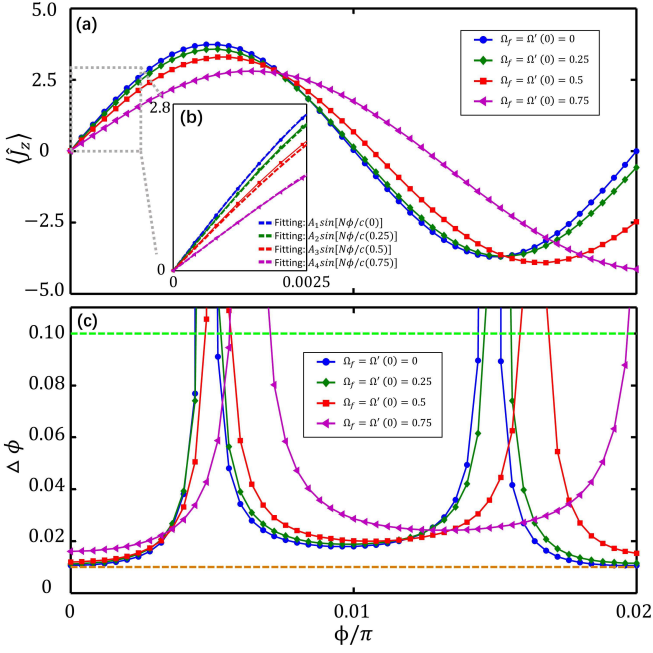}
\caption{\label{Fig2}(color online). Results of the Bose-Josephson model: (a) The final half population difference $\langle\hat J_z\rangle$ versus the accumulated phase $\phi$. (b) The inset shows the magnified region of $\langle \hat J_z\rangle$ near $\phi=0$, in which the dashed lines are fitted according to the sine function. Four different input states $|\Psi(\Omega_f)\rangle_G$, which correspond to $\Omega_f=(0, 0.25, 0.5, 0.75)$, are denoted by blue circles, green diamonds, red squares and purple triangles, respectively. In the recombination process, $\Omega'(t)$ is swept from $\Omega_f$ to the optimal value $\Omega_{opt}'$. (c) The phase measurement precision $\Delta\phi$ versus $\phi$. The green and orange dashed lines indicate the standard quantum limit and the Heisenberg limit, respectively.}
\end{figure}

\begin{figure}[!t]
\includegraphics[width=0.9\columnwidth]{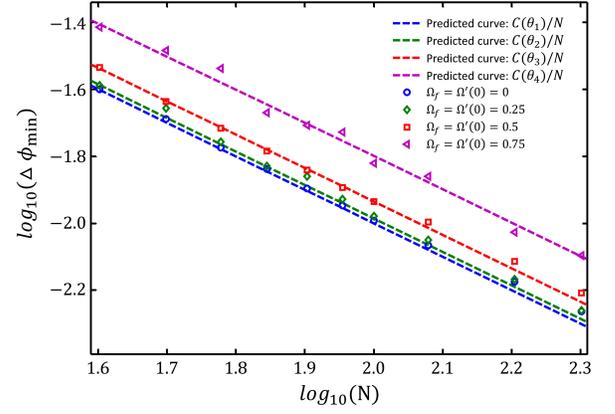}
\caption{\label{Fig3}(color online). The log-log precision scaling $\Delta\phi_{\text{min}}$ versus total atomic number $N$ for Bose-Josephson system via driving through QPTs with different input states $|\Psi(\Omega_f)\rangle_G$. The blue circles, green diamonds, red squares and purple triangles correspond to $\Omega_f=(0, 0.25, 0.5, 0.75)$, respectively. Their precisions follow the Heisenberg-limited scaling. The dashed lines are the predicted precision scaling obtained by approximating the input states $|\Psi(\Omega_f)\rangle_G$ as corresponding spin cat states (see Supplementary Material). }
\end{figure}

Through the interrogation process, the input state $|\Psi(\Omega_f)\rangle_G$ evolves into,
\begin{equation}\label{evo_state}
    |\Psi(\phi)\rangle= \hat U(\phi)|\Psi(\Omega_f)\rangle_G=e^{-i \phi \hat J_z}|\Psi(\Omega_f)\rangle_G,
\end{equation}
with the accumulated phase $\phi=\omega T$.
Here $\omega$ is the energy difference between $|a\rangle$ and $|b\rangle$ and $T$ denotes the interrogation time.
Then, after the interrogation process, the system undergoes another adiabatic process,
\begin{equation}\label{Omega}
\Omega'(t)=
\begin{cases}
\begin{split}
\Omega_f + \beta_2 t, \quad \quad \quad 0 \le t \le \tau_c',\\
\Omega_c + \beta_1 (t-\tau_c'),   \tau_c' < t \le \tau',
\end{split}
\end{cases}
\end{equation}
which is the inverse process of the beam splitting~\eqref{Omega1}.
That is, the Josephson coupling strength is swept from $\Omega'(0)=\Omega_f$ to $\Omega'(\tau_c')=\Omega_c$ with the sweeping rate $\beta_2$, and then from $\Omega_c$ to $\Omega'(\tau')=\Omega_{opt}'$ with the sweeping rate $\beta_1$.
In practice, it is unnecessary sweeping back to $\Omega_0$, since there exist some specific values $\Omega_{opt}'$ ($\Omega_c<\Omega_{opt}'<\Omega_0$) that can achieve the optimal phase estimation (see Supplementary Material).

When the Josephson coupling strength $\Omega'(t)$ is swept to the optimal point $\Omega_{opt}'$, we apply a $\pi\over2$-pulse and then measure the half population difference $\hat J_z$ to estimate the accumulated phase $\phi$.
%
%
%
In Fig.~\ref{Fig2}~(a), we show the expectations $\langle \hat J_z \rangle$ with respect to the accumulated phase $\phi$.
%
In our calculation, we choose $N=100$, $\Omega_0=11$, $\beta_1=0.1$ and $\beta_2=0.005$.
According to the error propagation formula, we obtain the measurement precision $\Delta\phi=\frac{\sqrt{\langle \hat J^2_z \rangle-\langle \hat J_z \rangle^2}}{|\partial \langle \hat J_z \rangle / \partial \phi|}$ versus $\phi$, see Fig.~\ref{Fig2}~(c).

The population measurement is efficient to estimate the accumulated phase $\phi$, especially near $\phi=0$.
For the input state $|\Psi(\Omega_f=0)\rangle_G$, the dependence of $\langle \hat J_z \rangle$ on $\phi$ is exactly proportional to $\sin(N\phi)$.
For $|\Psi(\Omega_f)\rangle_G$ with larger $\Omega_f$, the dependence of $\langle \hat J_z \rangle$ on $\phi$ gradually deviate the sinusoidal oscillation when $\phi$ increases. The oscillation frequency decreases with $\Omega_f$ and the amplitude is no longer fixed when $\phi$ is apart from $0$.
%
Nevertheless, near $\phi=0$, the expectation $\langle \hat J_z \rangle$ can still be approximated in sine function for most $|\Psi(\Omega_f)\rangle_G$, i.e., $\langle \hat J_z \rangle \simeq A(\Omega_f) \sin \left[N\phi/c(\Omega_f)\right]$, as shown in Fig.~\ref{Fig2}~(b).
Here, for $N=100$, $c(0)=1$, $c(0.25)=1.03$, $c(0.5)=1.16$ and $c(0.75)=1.58$, where the oscillation frequency $N/c(\Omega_f)$ decreases as $\Omega_f$.
The expectation $\langle \hat J_z^2 \rangle$ also oscillates with respect to $\phi$ and its minimum $B(\Omega_f)$ locates at $\phi=0$ for all $|\Psi(\Omega_f)\rangle_G$.
Thus, one can easily obtain the minimum phase uncertainty $\Delta\phi_{\text{min}}=\frac{\sqrt{B(\Omega_f)}c(\Omega_f)}{A(\Omega_f) N}$ for $|\Psi(\Omega_f)\rangle_G$ at $\phi=0$.
The minimum phase uncertainty at $\phi=0$ is inversely proportional to total atomic number $N$, which attains the Heisenberg limit.

To further confirm the Heisenberg scaling, we calculate the minimum phase uncertainty $\Delta\phi_{\text{min}}$ under different total particle numbers $N$, see Fig.~\ref{Fig3} for the log-log scaling of $\Delta\phi_{\text{min}}$ versus $N$.
For the input states $|\Psi(\Omega_f)\rangle_G$ with $\Omega_f=(0, 0.25, 0.5, 0.75)$, all their precisions follow the Heisenberg scaling but multiplied by different coefficients dependent on $\Omega_f$, i.e, $\Delta\phi_{\text{min}} \propto \tilde{C}(\Omega_f)/N$ (see Supplementary Material).
Assuming $N=100$ and $|\chi|/N \approx 0.5$ Hz~\cite{Gross2010, Riedel2010}, the duration for beam splitting can be estimated as $\tau\approx 6 $ s for $|\Psi(\Omega_f=0)\rangle_G$ and $\tau\approx 3 $ s for $|\Psi(\Omega_f=0.75)\rangle_G$.

\textit{Transverse-field Ising model.---}
Our scheme is also valid in a transverse-field Ising model, which can be realized with an array of ultracold ions~\cite{Porras2004, Friedenauer2008, Kim2010, Islam2011, Monz2011, Jurcevic2014, Jurcevic2017}.
The Hamiltonian reads~\cite{Elliott1970},
\begin{equation}\label{Ham_Ising}
    H_{T} = H_{I} + H_{B} = \sum_{i<j} J_{ij}\hat{\sigma}_z^i \hat{\sigma}_z^j- B\sum_{i} \hat{\sigma}_x^i,
\end{equation}
where $\hat{\sigma}_{x,z}^i$ are Pauli matrices for the $i$-th spin, $J_{ij}=J/|i-j|^3$ parameterizes the effective spin-spin interaction, and $B$ denotes an effective transverse magnetic field.
When $J=0$ (i.e. $H_B$ dominates), the system groundstate is a paramagnetic state of all spins aligned along the magnetic field.
If $B=0$ and $J<0$, the system groundstate is a superposition of two degenerated ferromagnetic states of all spins in either spin-up or spin-down.
The equal-probability superposition of these two degenerate groundstates is known as a GHZ state.

In the beam splitting process, the sweeping can be described as~\cite{Hu2012}
\begin{equation}\label{Sweep1}
\begin{cases}
\begin{split}
B(t) = B_0, \quad J(t)=2J_0 t/\tau, \quad \quad 0 \le t \le \tau/2,\\
B(t) = 2B_0(1-t/\tau), \quad J(t)=J_0,  \quad \tau/2 < t \le \tau,
\end{split}
\end{cases}
\end{equation}
where $B(0)=B_0>0$, $J(0)=0$, and the initial state is a spin coherent state $|\psi_I\rangle=\left[\frac{1}{\sqrt{2}}(|\uparrow\rangle+|\downarrow\rangle)\right]^{\otimes N}$.
In the first stage, the transverse field $B_0$ remain unchanged and the nearest-neighboring interaction is linear swept from $0$ to $J_0$.
In the second stage, the nearest-neighboring interaction is fixed to $J_0$ and the transverse field linearly decreases from $B_0$ to $0$.
If the sweeping is sufficiently slow ($\tau$ is large enough), the evolved state $|\psi_E\rangle$ would be very close to a GHZ state.

Given an input state, the phase accumulation process obeys,
\begin{equation}
H_{\omega}=\frac{\omega}{2} \sum_{i} \hat{\sigma}_{z}^{i}=\omega \hat M_{z},
\end{equation}
where the accumulated phase is given as $\phi=\omega T$ with $\omega$ the transition frequency and the interrogation time $T$.
Then, the inverse driving is applied on the output state for recombination.

The beam recombination process is described by
\begin{equation}\label{Sweep2}
\begin{cases}
\begin{split}
B(t) = 2B_0 t/\tau, \quad J(t)=J_0, \quad \quad 0 \le t \le \tau/2,\\
B(t) = B_0, \quad J(t)=2J_0(1- t/\tau),  \quad \tau/2 < t \le \tau',
\end{split}
\end{cases}
\end{equation}
where the nearest-neighboring interaction remains $J_0$ and the transverse field is linearly swept from $0$ to $B_0$ for $0 \le t \le \tau/2$, and then $B_0$ is fixed, the nearest-neighboring interaction is changed from $J_0$ towards $0$ for $\tau/2 < t \le \tau'$.

\begin{figure}[!t]
\includegraphics[width=\columnwidth]{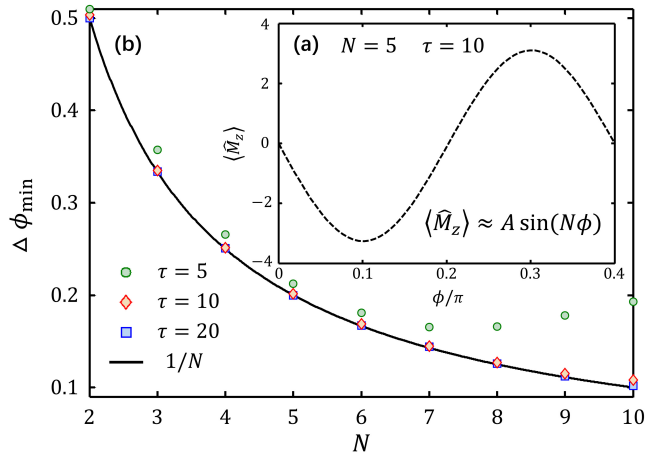}
\caption{\label{Fig4}(color online). Results of the transverse-field Ising model: (a) The final half population difference versus the accumulated phase $\phi$  for $N=5$ and $\tau=10$. (b) The phase precision versus the total particle number $N$ under different sweeping duration $\tau$. }
\end{figure}

The population measurement shows that, if the sweeping is sufficiently slow, the precision follows the Heisenberg scaling. In our calculations, we choose $B_0=1$ and $J_0=-1$.
Fig.~\ref{Fig4}~(a) shows the final half population difference for $N=5$ and $\tau=10$.
The half population difference is a sinusoidal function of the accumulated phase, $\langle \hat{M}_z \rangle \approx A \sin(N\phi)$, where $A$ is the amplitude related to $N$ and $\tau'$.
Thus one can extract the phase without single-particle resolved detectors.
Fig.~\ref{Fig4}~(b) shows the phase precision $\Delta\phi_{\text{min}}$ versus the total particle number $N$ for different sweeping duration $\tau$.
The precision follows the Heisenberg scaling when the sweeping is adiabatic (e.g., $\tau=20$).
When the sweeping is fast (e.g., $\tau=5$), the precision becomes more and more deviated from the ultimate bound as $N$ becomes larger and larger.
Obviously, $\tau=10$ is enough for Heisenberg-limited measurement.
Given $B_0 =J_0= 1$ kHz~\cite{Monz2011, Islam2011}, the duration $\tau=10$ corresponds to $10$ ms.

\textit{Discussion and Conclusion.---}
We have presented a scheme for precision metrology via driving through QPTs between non-degenerate and degenerate regimes.
Different from the scheme via QPTs between superfluid and Mott-insulator~\cite{Dunningham2002,Dunningham2004}, in which the degenerate regime is absent, our scheme uses the entangled non-Gaussian states for phase accumulation instead of the entangled Gaussian states.
In our scheme, adiabatic symmetry-breaking QPTs~\cite{Trenkwalder2016} are used to generate entangled non-Gaussian cat states~\cite{Lee2006, Lee2009, Huang2015}.
More importantly, due to the parity conservation~\cite{Xing2016}, the adiabatic processes are robust against excitations.
Thus, the sweeping rates in our state preparation and recombination can be achieved via  currently available experiment techniques.
Different from the scheme in~\cite{Dunningham2002}, in which two $\pi/2$ pulses sandwich the phase interrogation, our scheme does not need these pulses.

Our scheme paves a new way to utilize adiabatic QPTs for implementing high-precision interferometry with entangled non-Gaussian states.
In addition to taking the role of beam splitters, the adiabatic QPTs are used to entangle/disentangle particles at the same time.
Moreover, the beam recombination via reversed sweeping through QPTs does not bring any additional complexity in experiments, and the Heisenberg-limited measurement can be achieved without single-particle resolved detection.

\acknowledgements{J. Huang and M. Zhuang contribute equally to this work. This work is supported by the National Natural Science Foundation of China (NNSFC) under Grants No. 11374375, No. 11574405 and No. 11704420. J. H. is partially supported by National Postdoctoral Program for Innovative Talents of China (BX201600198).}

\end{document}